# Magneto-optical properties of a new group-IV ferromagnetic semiconductor $Ge_{1-x}Fe_x$ grown by low-temperature molecular beam epitaxy

Yusuke Shuto
Department of Electronic Engineering, The University of Tokyo, 7-3-1 Hongo, Bunkyo-ku, Tokyo 113-8656, Japan

Masaaki Tanaka
Department of Electronic Engineering, The University of Tokyo, 7-3-1 Hongo, Bunkyo-ku, Tokyo 113-8656, Japan, and SORST, Japan Science and Technology Agency, 4-1-8 Honcho, Kawaguchi, Saitama 332-0012, Japan

Satoshi Sugahara
Department of Frontier Informatics, The University of Tokyo, 5-1-5 Kashiwanoha, Kashiwa-shi, Chiba 277-8583, Japan

A new group-IV ferromagnetic semiconductor, $Ge_{1-x}Fe_x$, was successfully grown by low-temperature molecular beam epitaxy (LT-MBE) without precipitation of ferromagnetic Ge-Fe intermetallic compounds. The ferromagnetism of $Ge_{1-x}Fe_x$ films was investigated by magnetic circular dichroism (MCD). In particular, the influence of the Fe content ($F_{Fe}/F_{Ge}$ =1 - 10%) and growth temperature (100, 200$^O$C) on the ferromagnetism was carefully studied. The MCD measurements revealed that the band structure of the $Ge_{1-x}Fe_x$ films was identical with that of bulk Ge, and that the large spin splitting of the band structure was induced by the incorporation of Fe atoms into the Ge matrix, indicating the existence of s,p-d exchange interactions. The $Ge_{1-x}Fe_x$ films showed ferromagnetic behavior and the ferromagnetic transition temperature linearly increased with increasing the Fe composition. These results indicate that the epitaxially grown $Ge_{1-x}Fe_x$ is an intrinsic ferromagnetic semiconductor.

Recently, ferromagnetic semiconductors have attracted a great deal of attention, since they are expected to be used for semiconductor-based spintronic devices that have dual advantages of information processing based on carrier transport and information storage based on nonvolatile magnetization. In addition, the carrier-induced ferromagnetism of ferromagnetic semiconductors can lead to very useful features to manipulate the magnetization or carrier spin, such as field-effect-control of the magnetization. Thus, ferromagnetic semiconductors are expected to be utilized to produce more functional devices, compared with ordinary ferromagnetic-metal-based spintronics devices. Although compound-semiconductor-based ferromagnetic semiconductors, such as (In,Mn)As [1, 2], (Ga,Mn)As [3, 4, 5], and (Zn,Cr)Te [6, 7] have been widely studied so far, new ferromagnetic semiconductors that are compatible with present Si-technology are strongly needed in order to develop integrated electronics employing spin degrees of freedom. In particular, Si- and Ge-based ferromagnetic semiconductors are expected to be highly attractive for novel integrated spintronics based on Si-technology [8].

The epitaxial growth of Mn-doped Ge ($Ge_{1-x}Mn_x$) films was reported by Park *et al.* [9]. They also demonstrated the anomalous Hall effect with carrier-induced magnetization control of the $Ge_{1-x}Mn_x$ films by applying gate voltages. There are also several reports that focused on Ge-based ferromagnetic semiconductors [10-12], including high Curie temperatures close to room temperature [13-15]. Nevertheless, owing to the lack of direct evidence for the spin splitting of their band structure induced by s,p-d exchange interactions, the origin of their ferromagnetism has been still under investigation [16].

In this paper, we present the epitaxial growth and ferromagnetism of new group IV

ferromagnetic semiconductor $Ge_{1-x}Fe_x$. In order to clarify the origin of the ferromagnetism in $Ge_{1-x}Fe_x$, magnetic circular dichroism (MCD) was applied. It was found that the spin splitting of the band structure of epitaxially grown $Ge_{1-x}Fe_x$ films originates from the s,p-d exchange interactions that are induced by the incorporation of Fe atoms into the Ge matrix, indicating that $Ge_{1-x}Fe_x$ is an intrinsic ferromagnetic semiconductor.

$Ge_{1-x}Fe_x$ films were epitaxially grown on Ge(001) substrates by low temperature molecular beam epitaxy (LT-MBE). A Ge(001) substrate was introduced into the MBE growth chamber using an oil-free load-lock system, in order to eliminate any carbon-related contamination. After degassing the sample at a substrate temperature ($T_S$) of 400$^O$C for 30 min and successive thermal cleaning at $T_S$ =700$^O$C for 15 min, an 8-nm-thick Ge buffer layer was grown at $T_S$ =100$^O$C, followed by the growth of a 16-nm-thick $Ge_{1-x}Fe_x$ film at $T_S$ =100 or 200$^O$C. The Ge flux (measured by an ion gauge) was fixed at 2.4 x 10$^{-6}$ Pa and the Fe concentration of the $Ge_{1-x}Fe_x$ film was controlled by the Fe and Ge flux ratio ($F_{Fe}/F_{Ge}$). The $F_{Fe}/F_{Ge}$ value was varied from 1 to 10%. *In-situ* reflection high-energy electron diffraction (RHEED) was used to monitor the crystallinity and morphology during the growth. Although the diffraction pattern of the Ge buffer layer surface showed intense and sharp 2X2 streaks, the pattern was quickly changed depending on the $F_{Fe}/F_{Ge}$ ratio at the initial stage of the growth of the $Ge_{1-x}Fe_x$ layer; intense and sharp 2X2 streaks for $F_{Fe}/F_{Ge}$ $\leq$3%, weak and diffuse 2X2 streaks for $F_{Fe}/F_{Ge}$ =5%, and 1X1 spots for $F_{Fe}/F_{Ge}$ >5%. Throughout the MBE growth, the RHEED pattern maintained diamond crystal structure, and no extra spots were observed for $F_{Fe}/F_{Ge}$ $\leq$10% (that is the highest flux ratio examined here), indicating that the precipitation of other crystal phases was not induced. Thus, the $Ge_{1-x}Fe_x$ film with the diamond crystal structure was epitaxially grown up to $F_{Fe}/F_{Ge}$ =10%, although the crystallinity was deteriorated with increasing $F_{Fe}/F_{Ge}$.

The crystal structure was characterized by cross-sectional high-resolution transmission electron microscopy (HRTEM). We observed two HRTEM lattice images of an epitaxial $Ge_{1-x}Fe_x$ film with $F_{Fe}/F_{Ge}$ =5% grown at $T_S$ =200$^O$C, with the projection directions along the exact [110] axis and slightly tilted from the [110] axis. It is well recognized that when the projection direction in TEM observation is slightly tilted from the zone axis, the contrast due to the difference of the chemical composition is enhanced. The lattice image of the epitaxial layer taken with the exact [110] axis projection was homogeneous without contrast variation. The $Ge_{1-x}Fe_x$ layer is epitaxially aligned on the Ge buffer layer with the diamond crystal structure and without threading dislocations. The surface of the films was atomically smooth with roughness of less than 1nm, which agreed with the RHEED observations, as previously described. When the TEM image was taken with slightly tilted angle from the exact [110] axis, cluster-like dark regions were observed in the epitaxial layer. Local area energy dispersive X-ray spectroscopy (EDX) measurements revealed that the Fe atoms were nonuniformly distributed in the epitaxial layer and these dark regions were caused by the higher density of the Fe atoms. The Fe concentrations of the dark regions and the surrounding matrix region were ~12% and ~4%, respectively. However, it was confirmed from transmission electron diffraction (TED) measurements that these dark regions also showed diamond structure as well as the matrix, and any other crystal structures were not detected, although twin defects were observed in the dark regions. The detailed analysis of the TEM observation will be reported elsewhere.

Figure 1 shows MCD spectra of $Ge_{1-x}Fe_x$ films with $F_{Fe}/F_{Ge}$ =1 and 5% grown at $T_S$ =100$^O$C, where the samples were measured at 10K with reflection configuration. A magnetic field was applied perpendicular to the film plane. The MCD spectrum of bulk Ge is also shown as a reference. The spectral features (i.e., MCD shape around the $E_1$ and $E_0$' transition energies) of the $Ge_{1-x}Fe_x$ film with $F_{Fe}/F_{Ge}$ =1% were very similar to those of bulk Ge. The MCD intensity corresponding to the $E_1$ transition energy of bulk Ge was significantly enhanced, although the peak energy was slightly shifted to higher energy. The MCD intensity at the $E_1$ critical point was much enhanced for the $Ge_{1-x}Fe_x$ film with $F_{Fe}/F_{Ge}$ =5%. This means that the spin splitting of the band structure is induced by the incorporation of Fe atoms and implies the existence of s,p-d exchange interactions [17]. Offset-like MCD signals appeared in the whole energy range examined here (1.5 – 4.5eV) for the $Ge_{1-x}Fe_x$ films with $F_{Fe}/F_{Ge}$ $\geq$3%. Although the origin of this offset is not clear at this stage, this is not caused by ferromagnetic precipitates as discussed later. Note that the periodic peaks above 3.2eV for the sample with $F_{Fe}/F_{Ge}$ =5% grown at $T_S$ =100$^O$C was due to the extrinsic interference of an adsorbed moisture layer, since the residual moisture in our vacuum MCD system was adsorbed on the surface of the $Ge_{1-x}Fe_x$ films during measurements. In the $Ge_{1-x}Fe_x$ films with $F_{Fe}/F_{Ge}$ $\geq$3%, the MCD peak related to the $E_0$' transition energy was not clearly distinguished due to the offset MCD signals and the extrinsic interference. When the $Ge_{1-x}Fe_x$ films were grown at $T_S$ =200$^O$C, the spectral features of MCD measured at low-temperatures were qualitatively the same as that of the $Ge_{1-x}Fe_x$ films grown at $T_S$ =100$^O$C, as also shown in Fig. 1. However, the ferromagnetic transition temperatures strongly depend on $T_S$ as discussed below.

Figure 2 shows the magnetic field dependence of the MCD intensities (hysteresis loops) at several photon energies ($E_1$ energy (2.34eV), near $E_0$' (2.88eV), and other points (4.13 and 1.91eV) on the offset-signal region) for the $Ge_{1-x}Fe_x$ film with $F_{Fe}/F_{Ge}$ =5% grown at $T_S$ =200$^O$C. Normalized intensities of these MCD signals are also shown in the inset, in which the data were normalized by the MCD intensities at 1T. The MCD signals at these photon energies exhibited clear ferromagnetic hysteresis loops, and the normalized loops are completely identical with each other. This indicates that the ferromagnetism of the $Ge_{1-x}Fe_x$ film comes from a single phase origin, and the formation of ferromagnetic precipitates can be excluded. Therefore, the MCD measurements shown in Fig. 2 indicate that the epitaxially grown $Ge_{1-x}Fe_x$ films are magnetically homogeneous, which are also consistent with the TEM observation described previously.

The ferromagnetic transition temperature $T_C$ was estimated from Arrott plots, i.e., (MCD)$^2$ - B/MCD relation. Figure 3 shows the Arrott plots obtained from the magnetic field dependence of the MCD intensity at the $E_1$ critical point (2.34eV) for the $Ge_{1-x}Fe_x$ film with $F_{Fe}/F_{Ge}$ =5% grown at $T_S$ =200$^O$C. The estimated $T_C$ value was 90 - 100K. Figure 4 shows the estimated $T_C$ as a function of $F_{Fe}/F_{Ge}$ for the samples grown at $T_S$ =200$^O$C. $T_C$ increased linearly with $F_{Fe}/F_{Ge}$. When the samples were grown at $T_S$ =100$^O$C, $T_C$ was reduced. For example, the $T_C$ value of the $Ge_{1-x}Fe_x$ film with $F_{Fe}/F_{Ge}$ =5% grown at $T_S$ =100$^O$C was ~40K. The highest $T_C$ was 170K for the $Ge_{1-x}Fe_x$ film with $F_{Fe}/F_{Ge}$ =10% grown at $T_S$ =200$^O$C.

In summary, new group-IV ferromagnetic semiconductor $Ge_{1-x}Fe_x$ was epitaxially grown on Ge(001) substrates by LT-MBE. MCD measurements revealed that the $Ge_{1-x}Fe_x$ films exhibit ferromagnetic ordering and their spectral features certainly originate from the band structure of Ge with largely enhanced spin splitting of the band edge induced by the s,p-d exchange interactions. These results indicate that the epitaxially grown $Ge_{1-x}Fe_x$ film is an "intrinsic" ferromagnetic semiconductor.

8**Acknowledgements**

This work was partly supported by Industrial Technology Research Grant Program of NEDO, PRESTO and SORST Programs of JST, Grant-in-Aids for Scientific Research, IT Program of RR2002 of MEXT, and Toray Science Foundation.

**References**

[1] H. Munekata, H. Ohno, S. von Molnar, A. Segmuller, L. L. Chang, and L. Esaki, Phys. Rev. Lett. **63**, 1849 (1989).

[2] H. Ohno, H. Munekata, T. Penny, S. von Molnar, and L. L. Chang, Phys. Rev. Lett. **68**, 2664 (1992).

[3] H. Ohno, A. Shen, F. Matsukura, A. Oiwa, A. Endo, S. Katsumoto, and H. Iye, Appl. Phys. Lett. **69**, 363 (1996).

[4] T. Hayashi, M. Tanaka, T. Nishinaga, H. Shimada, H. Tsuchiya, and Y. Otuka, J. Crystal Growth **175/176**, 1063 (1997)

[5] K. Y. Wang, R. P. Campion, K. W. Edmonds, M. Sawicki, T. Dietl, C. T. Foxon, and B. L. Gallagher, cond-mat/0411475.

[6] W. Mac, A. Twardowski, and M. Demianiuk, Phys. Rev. B **54**, 5528 (1996).

[7] H. Saito, V. Zayets, S. Yamagata, and K. Ando, Phys. Rev. Lett. **90**, 207202 (2003).

[8] S. Sugahara and M.Tanaka, J. Appl. Phys. **97**,10D503 (2005).

[9] Y. D. Park, A. T. Hanbicki, S. C. Erwin, C. S. Hellberg, J. M. Sullivan, J. E. Mattson, T. F. Ambrose, A. Wilson, G. Spanos, and B. T. Jonker, Science **295**, 651 (2002).

[10] G. Kioseoglou, A. T. Hanbicki, C. H. Li, S. C. Erwin, R. Goswami, and B. T. Jonker, Appl. Phys. Lett. **84**, 1725 (2004)

[11] F. Tsui, L. He, A. Tkachuk, S. Vogt, and Y. S. Chu, Phy. Rev. B **69**, 081304(R) (2004).

[12] A. P. Li, J. Shen, J. R. Thompson, and H. H. Weitering, Appl. Phys. Lett. **86**, 152507 (2005).

[13] N. Pinto, L. Morresi, M. Ficcadenti, R. Murri, F. D'Orazio, F. Lucari, L. Boarino, and G. Amato, Phys. Rev. B **72**, 165203 (2005).

[14] J. -S. Kang, G. Kim, S. C. Wi, S. S. Lee, S. Choi, Sunglae Cho, S. W. Han, K. H. Kim, H. J. Song, H. J. Shin, A. Sekiyama, S. Kasai, S. Suga, and B. I. Min, Phys. Rev. Lett. **94**, 147202 (2005).

[15] G. Kim, S. C. Wi, S. S. Lee, J. –S. Kang, S. Y. Choi, Sunglae Cho, S. W. Han, K. H. Kim, H. J. Song, and H. J. Shin, J. Appl. Phys. **97**, 10A307 (2005).

[16] S. Sugahara, K. L. Lee, S. Yada, and M. Tanaka, Jpn. J. Appl. Phys. Submitted.

[17] J. K. Furdyna, J. Appl. Phys. **64**, R29 (1988)

FIG. 1. MCD spectra of the $Ge_{1-x}Fe_x$ thin films with $F_{Fe}/F_{Ge}$ =1 and 5% grown at substrate temperatures ($T_S$) of 100 and 200°C, and that of bulk Ge as a reference. The MCD measurements were carried out at 10K with a reflection configuration. A magnetic field of 1T was applied perpendicular to the film plane.

FIG. 2. Magnetic field dependence of the MCD intensity (hysteresis loops) at several photon energy points measured at 10K for the $Ge_{1-x}Fe_x$ film with $F_{Fe}/F_{Ge}$ =5% and $T_S$ =200°C. The inset shows normalized hysteresis loops, in which the data were normalized by the MCD intensities at 1T.

FIG. 3. Arrott plots of the MCD hysteresis loops at the $E_1$ critical point measured at various temperatures for the $Ge_{1-x}Fe_x$ film with $F_{Fe}/F_{Ge}$ =5% grown at $T_S$ =200°C. The $T_C$ value of this sample was estimated to be 90 - 100K.

Fig. 4. $T_C$ as a function of $F_{Fe}/F_{Ge}$, estimated from the Arrott plots for the $Ge_{1-x}Fe_x$ films grown at $T_S$=200°C.

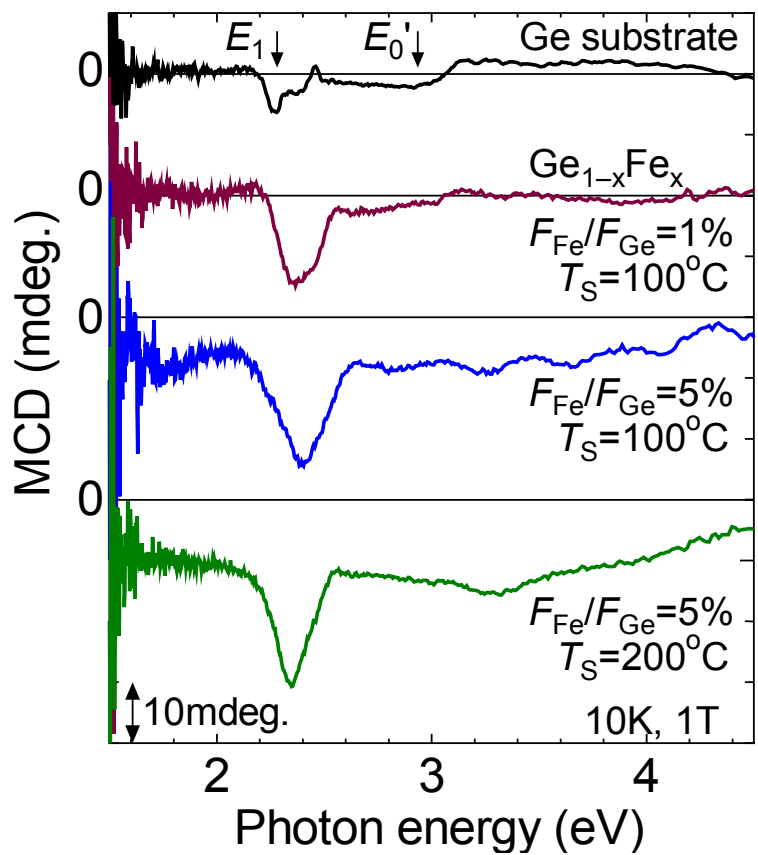

Fig. 1.  Shuto, *et al.*

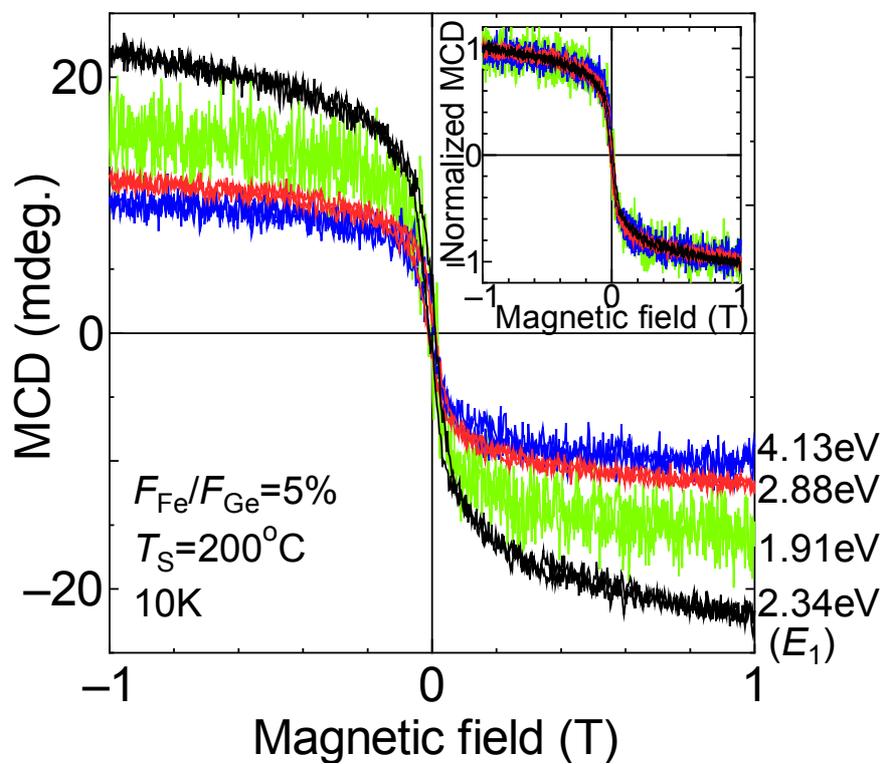

Fig. 2. Shuto, *et al.*

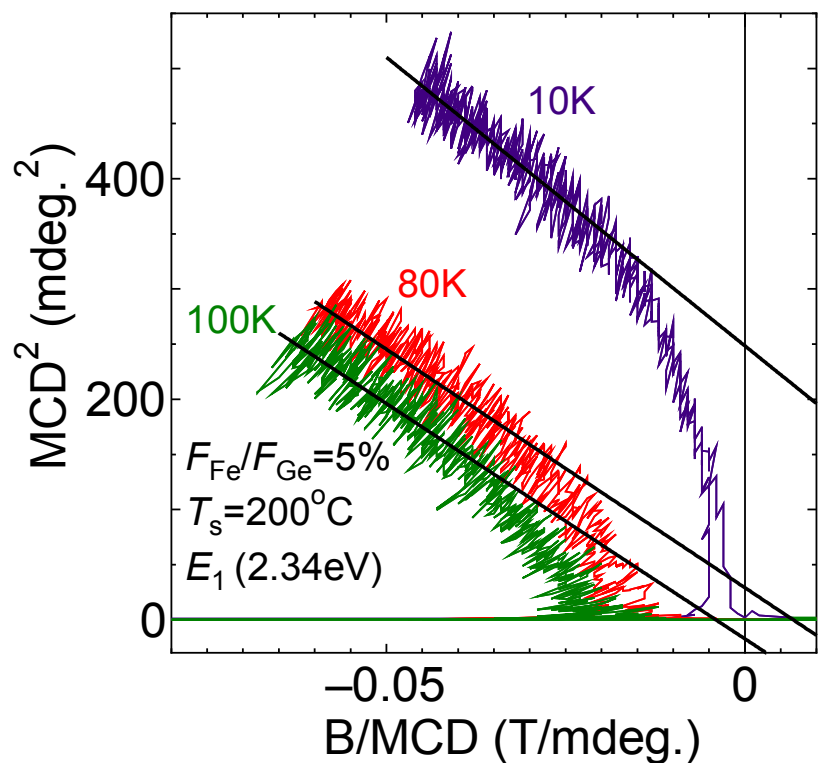

Fig. 3. Shuto, *et al.*

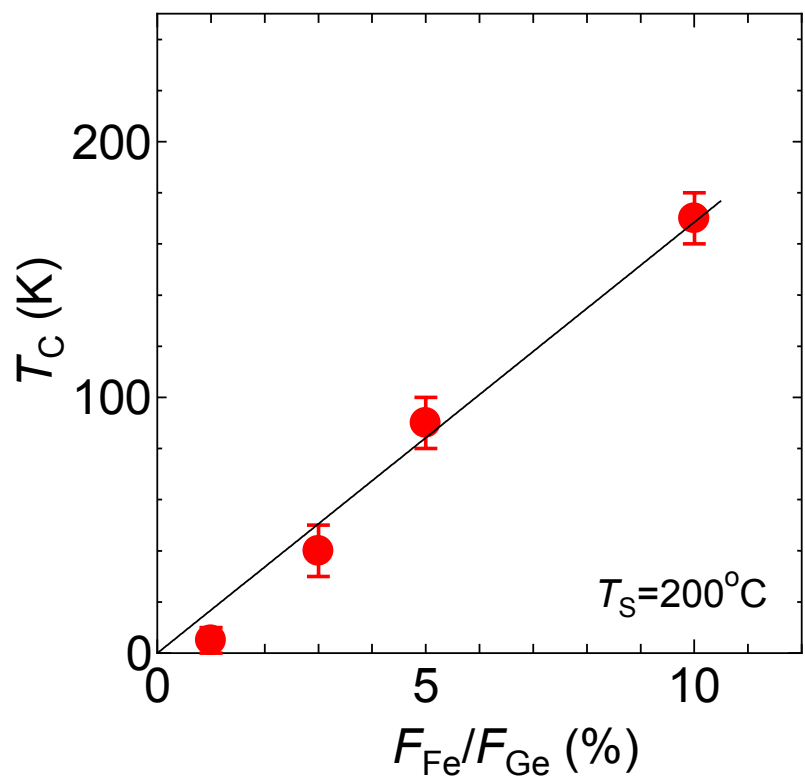

Fig. 4.  Shuto, *et al.*